\documentclass[11pt]{article}
\usepackage{latexsym}
\newcommand{\be}{\begin{equation}}
\newcommand{\ee}{\end{equation}\noindent}
\newcommand{\bear}{\begin{eqnarray}}
\newcommand{\ear}{\end{eqnarray}\noindent}
\newcommand{\no}{\noindent}
\date{}
\renewcommand{\theequation}{\arabic{section}.\arabic{equation}}

\newcommand{\slD}{\raise.15ex\hbox{$/$}\kern-.57em\hbox{$D$}}
\newcommand{\slpartial}{\raise.15ex\hbox{$/$}\kern-.57em\hbox{$\partial$}}
\newcommand{\slG}{{{\dot G}\!\!\!\! \raise.15ex\hbox {/}}}

\def\a{\alpha}\def\adt{\dot\alpha}
\def\th{\theta}
\def\non{\nonumber}
\def\beqn*{\begin{eqnarray*}}
\def\eqn*{\end{eqnarray*}}

\def\square{\kern1pt\vbox{\hrule height 1.2pt\hbox{\vrule width 1.2pt
   \hskip 3pt\vbox{\vskip 6pt}\hskip 3pt\vrule width 0.6pt}
   \hrule height 0.6pt}\kern1pt}
\def\slash#1{#1\!\!\!\raise.15ex\hbox {/}}

\def\dps{\displaystyle}

\def\4piTD{{(4\pi T)}^{-{D\over 2}}}
\def\4piT4{{(4\pi T)}^{-2}}

\def\Tintm4{{\dps\int_{0}^{\infty}}{dT\over T}\,e^{-m^2T}
    {(4\pi T)}^{-2}}
\def\Tintm{{\dps\int_{0}^{\infty}}{dT\over T}\,e^{-m^2T}}

\def\tr{{\rm tr}\,}

\def\be{\begin{equation}}\def\ee{\end{equation}}
\def\bea{\begin{eqnarray}}\def\eea{\end{eqnarray}}
\def\ba{\begin{array}}\def\ea{\end{array}}
\def\bea{\begin{eqnarray}}\def\barr{\begin{array}}\def\earr{\end{array}}
\def\eea{\end{eqnarray}}
\begin{document} \input feynman
\newcommand{\ho}[1]{$\, ^{#1}$}
\newcommand{\hoch}[1]{$\, ^{#1}$}
\newcommand{\kings}{\it\small Department of Mathematics, King's College, London, UK}
\newcommand{\lapp}{\it\small LAPP, Annecy, France}
%%%%%%%%%%%%%%%%%%%%%%%%%%%%%%%%%%%%%%%%%%%%%%%%%%%%%%%%%%%%%%%%%%%%%%%%%%%%%

\newcommand{\auth}{\large B. Eden\hoch{a}, P.S. Howe\hoch{a},
C. Schubert\hoch{b}, E. Sokatchev\hoch{b} and P.C. West\hoch{a}}

\thispagestyle{empty}

%\begin{document}

\hfill{KCL-MTH-99-22}

\hfill{LAPTH-735/99}

\hfill{hep-th/???????} 

\hfill{\today}

\vspace{20pt}

\begin{center}
{\Large{\bf Simplifications of four-point functions in $N=4$ 
supersymmetric Yang-Mills theory at two loops}} \vspace{30pt} 

\auth

\vspace{15pt}

\begin{itemize}
\item [$^a$] \kings
\item [$^b$] {\it\small Laboratoire d'Annecy-le-Vieux de Physique
Th\'eorique\footnote{URA 1436 associ\'ee \`a l'Universit\'e de Savoie} LAPTH,
Chemin de Bellevue, B.P. 110, F-74941 Annecy-le-Vieux, France}
\end{itemize}

\vspace{60pt}

{\bf Abstract}

\end{center}

The superconformal Ward identities combined with $N=2$ harmonic 
analyticity are used to evaluate two-loop four-point correlation 
functions of gauge-invariant operators in $D=4,\ N=4$ 
supersymmetric Yang-Mills theory in terms of the well-known 
one-loop box integral. The result is confirmed by a direct 
numerical computation. \vfill 

\pagebreak
\renewcommand{\theequation}{\arabic{equation}}
\setcounter{equation}{0}

Supersymmetric Yang-Mills (SYM) theory with maximally extended 
($N=4$) supersymmetry in four dimensions has long been known to 
have some very interesting properties, in particular, it is 
ultra-violet finite and hence superconformally invariant quantum 
mechanically \cite{n4}. It also admits monopole solutions which 
fall into spin one multiplets \cite{os} and which play a crucial 
r\^ole in Montonen-Olive duality \cite{mo,sen}. There has been a 
great deal of recent interest in this theory as a superconformal 
field theory stimulated by the Maldacena conjecture via which it 
is related, in a certain limit, to IIB supergravity on 
$AdS_5\times S^5$ \cite{mal}. In a previous paper \cite{ehssw} we 
calculated four-point functions of certain gauge-invariant 
operators in this theory at two loops in perturbation theory using 
$N=2$ harmonic superspace Feynman diagrams \cite{harmgraph}. The 
result involves three functions, $A_1,\ A_2$ and $A_3$ which can 
be expressed in terms of differential operators acting on a scalar 
two-loop integral \footnote{The coefficient $A_2$ was 
independently calculated using $N=1$ superfields in 
\cite{gopasc}.}. The first two of these can be reduced quite 
straightforwardly to the well-known one-loop box integral but the 
third is much more complicated and in \cite{ehssw} we were unable 
to simplify it. As a consequence, we could not confirm the 
conformal invariance of the two-loop contribution to the 
four-point function. Furthermore, all three functions are needed 
to construct the full $N=4$ amplitude. In this note we show how 
this problem can be solved by using the superconformal Ward 
identities for harmonic-analytic operators in harmonic superspace. 
More precisely, we show that from the Ward identities combined 
with the dynamical requirement of harmonic analyticity follow 
differential constraints on the coefficient functions. Thus, $A_3$ 
is determined in terms of $A_1$ and $A_2$, so that it too reduces 
to a one-loop integral. The relation between the two- and one-loop 
integrals we have found is far from obvious, so in order to 
confirm it we have also performed a direct numerical computation. 

In $N=2$ harmonic superspace $N=4$ Yang-Mills theory is formulated 
in terms of an $N=2$ SYM multiplet, described by a prepotential 
$V^{++}$, and a hypermultiplet, $q^{+}$, both of which take their 
values in the Lie algebra of the gauge group $SU(N_c)$. Both of 
these superfields are Grassmann-analytic (G-analytic), that is 
they depend on half of the odd coordinates. We recall that $N=2$ 
harmonic superspace \cite{harmsup} is the product of Minkowski 
superspace $M$ with the sphere $S^2\sim SU(2)/U(1)$, and that the 
harmonic fields are defined on $M\times SU(2)$ but are equivariant 
with respect to the action of the $U(1)$ isotropy group of the 
sphere. The coordinates of this superspace are 
$\{x^{\a\adt},\th^{\a}_i,\bar\th^{\adt i},u_i^+, u_i^-\}$, where 
the $u^\pm_i$ together make up an $SU(2)$ matrix. A superfield 
$F^q(x,\th,u)$ which has $U(1)$ charge $q$ is called Grassmann 
analytic if it satisfies 

\be
D_{\a}^+ F^q:=u_i^+ D_{\a}^i F^q=0,\qquad \bar D_{\adt}^+ 
F^q:=u_i^+  \bar D_{\adt}^i F^q=0. \ee These constraints are 
easily solved in an appropriate superspace frame and amount to 
functional dependence on $\th^+:=u_i^+\th^i,\ 
\bar\th^+:=u_i^+\bar\th^i$ only (but not on $\th^-,\bar\th^-$). 
Thus, G-analyticity is just a kinematical constraint (similar to 
chirality in $N=1$ superspace). 

On shell, the hypermultiplet superfield $q^+$ has the dynamical 
property of harmonic analyticity (H-analyticity) 

\be\label{eqmot} D^{++} q^+ = 0, \ee in addition to the 
kinematical property of Grassmann analyticity. Here $D^{++},\ 
D^{--}= \overline{D^{++}}$ and $D_o$ are the standard derivatives 
on $SU(2)$ (left-invariant vector fields) which fulfill the 
algebraic relations of the associated Lie algebra. $D^{++}$ 
essentially corresponds to $\bar\partial$ on the sphere in the 
sense that a field of positive charge $q$ which is annihilated by 
$D^{++}$ corresponds precisely to a holomorphic section of the 
line bundle ${\cal O}(q)$ over $S^2$; its dependence on the 
harmonic variables is thereby restricted to be polynomial. In 
terms of component fields, the superfield equation (\ref{eqmot}) 
eliminates the infinite tower of auxiliary fields originating from 
the harmonic expansion on $S^2$. In addition, it puts the 
remaining physical fields on shell, therefore it is the field 
equation of the $N=2$ hypermultiplet. 

The operators we are interested in are the gauge-invariant 
bilinear combinations of the hypermultiplet, $\tr(q^+ q^+)$, 
$\tr(\widetilde q^+ q^+)$ and $\tr(\widetilde q^+ \widetilde 
q^+)$, where tilde denotes a conjugation which combines complex 
conjugation with the antipodal map on the sphere. These operators 
arise as ``components'' of the $N=4$ supercurrent multiplet of 
operators. Precisely this multiplet corresponds to $D=5,\ N=8$ 
supergravity on an $AdS$ background in the Maldacena conjecture, 
when it is decomposed into $N=2$ superfields.  In \cite{ehssw} we 
showed how it is possible to reconstruct the leading scalar term 
of the $N=4$ supercurrent four-point function from the leading 
terms of three $N=2$ hypermultiplet bilinear four-point functions. 
Given this leading term, the entire superfunction can be obtained 
by means of $N=4$ superconformal symmetry. 

The hypermultiplet bilinears satisfy both the G- and H-analyticity 
conditions on shell. Provided, therefore, that we are allowed to 
use the equations of motion of the underlying hypermultiplets, 
$n$-point Green's functions of such operators will be both G- and 
H-analytic functions on the product of $n$ copies of harmonic 
superspace. This circumstance, combined with the requirement of 
superconformal invariance of the $N=4$ theory, yields constraints 
on the Green's functions which can be understood as arising from 
the fact that analytic superconformal invariants have poles on 
$S^2$ while the Green's functions themselves must be regular. It 
is precisely these constraints which will allow us to establish 
the relation among the three coefficient functions of the 
four-point amplitude. 

We now summarise the results of reference \cite{ehssw} for the 
four-point function $\langle \widetilde q^+\widetilde q^+\vert 
q^+q^+\vert\widetilde q^+\widetilde q^+\vert q^+q^+\rangle $ (as 
explained in \cite{ehssw}, the other possible combinations of 
hypermultiplet bilinears can either be obtained from this one by 
permutations or vanish). Its leading (scalar) term in an expansion 
in the odd coordinates (denoted by ``$\vert_{\th=0}$'') is given 
by 

\bea\label{answer} 
 \langle \widetilde q^+\widetilde q^+ \vert &\hspace{-7cm}q^+q^+\vert 
  \widetilde q^+\widetilde q^+\vert
q^+q^+\rangle\mid_{\th=0} & \nonumber \\ &= \left[ (14)^2(23)^2 
A_1 + (12)^2(34)^2 A_2 + (12)(23)(34)(41) A_3\right]\;, \eea where 

\be\label{A12} A_1={(\partial_1+\partial_2)^2 f(1,2,3,4)\over 
x^2_{14}x^2_{23}}\;, \qquad A_2 ={(\partial_1+\partial_4)^2 
f(1,4,2,3)\over x^2_{12}x^2_{34}}\;, \ee and 

$$
A_3 = 4i\pi^2 {(x^2_{24} - x^2_{12} -x^2_{14}) g_3 + (x^2_{13} - x^2_{12}
-x^2_{23}) g_4 + (x^2_{24} - x^2_{23} -x^2_{34}) g_1 + (x^2_{13} - x^2_{14}
-x^2_{34}) g_2
\over x^2_{12}x^2_{14}x^2_{23}x^2_{34}}
$$

\be\label{A3}
 + {(\partial_2+\partial_3)^2 f(1,2,3,4)\over x^2_{14}x^2_{23}} +
{(\partial_1+\partial_2)^2 f(1,4,2,3)\over x^2_{12}x^2_{34}}\;.
\ee Here 

\be\label{2loop} f(x_1,x_2,x_3,x_4) = \int{d^4x_5 d^4x_6\over 
x^2_{15}x^2_{25}x^2_{36}x^2_{46}x^2_{56}} \ee and 

\be\label{1loop} g_4(x_1,x_2,x_3) = \int{d^4x_5\over 
x^2_{15}x^2_{25}x^2_{35}} \ee (and similarly for $g_1,\ g_2$ and 
$g_3$) are two- and one-loop scalar integrals, correspondingly. As 
usual $x_{ij}:=x_i-x_j$. The dependence of the Green's functions 
on the harmonic variables is polynomial (a corollary of 
H-analyticity) and is given by factors such as $(14)^2(23)^2$ 
where the abbreviation 

\be
(12):= \epsilon^{ji} u_i^+(1) u_j^+(2) \ee has been employed. 

The one-loop triangle integral $g_4$ (\ref{1loop}) is well-known 
\cite{thovel}. Following \cite{davussladder} we express it as 

\be\label{gtophi} g_4(x_1,x_2,x_3) = - {i\pi^2\over x_{12}^2} 
\Phi^{(1)} \Bigl( {x_{23}^2\over x_{12}^2} , {x_{31}^2\over 
x_{12}^2} \Bigr)\; \ee where 

\bear
\Phi^{(1)}(x,y)
=
{1\over \lambda}
\Biggl\lbrace
2\Bigl({\rm Li}_2(-\rho x)
+
{\rm Li}_2(-\rho y)\Bigr)
+\ln
{y\over x}
\ln
{{1+\rho y}\over {1+\rho x}}
+
\ln (\rho x)\ln (\rho y)
+
{\pi^2\over 3}
\Biggr\rbrace
\non\\
\label{Phi1explicit}
\ear\no
and

\bear \lambda(x,y)\equiv \sqrt{(1-x-y)^2-4xy}\ , \qquad 
\rho(x,y)\equiv 2(1-x-y+\lambda)^{-1}\ , \label{deflambdarho} 
\ear\no ${\rm Li}_2$ denoting the Euler dilogarithm. The two-loop 
scalar integral function $f(x_1,x_2,x_3,x_4)$ is  not yet known in 
explicit form. However, this is no impediment to the calculation 
of $A_1$ and $A_2$ since the combination of derivatives on $f$ 
appearing here can be straightforwardly reduced to a one-loop box 
integral by using translation invariance for one of the 
subintegrations. Doing this one finds \cite{ehssw} 

\bear A_1 &=& {4\pi^4\over x_{14}^2x_{23}^2x_{12}^2x_{34}^2} 
\Phi^{(1)} \Bigl( {x_{14}^2x_{23}^2\over x_{12}^2x_{34}^2} , 
{x_{13}^2x_{24}^2\over x_{12}^2x_{34}^2} \Bigr)\ . \label{resA1} 
\ear $A_2$ differs from $A_1$ only by the interchange 
$2\leftrightarrow 4$. In contrast to the evaluation of $A_1$ and 
$A_2$ the evaluation of $A_3$ is a formidable problem; the only 
explicit information on $A_3$ obtained in \cite{ehssw} was the 
cancellation of the leading ${1\over x_{14}^2}$ - pole in $A_3$ in 
the limit $x_{14}\to 0$. 

In the following we shall use conformal supersymmetry and harmonic
analyticity to obtain a constraint equation between the three 
coefficient functions $A_1,A_2,A_3$. By conformal invariance we 
can rewrite these in terms of conformal cross ratios as follows: 

\bear A_1 &=& {a(s,t)\over x_{14}^4x_{23}^4}\ , \non\\ A_2 &=& 
{b(s,t)\over x_{12}^4x_{34}^4}\ , \non\\ A_3 &=& {c(s,t)\over 
x_{12}^2x_{23}^2x_{34}^2x_{14}^2} \non\\ \label{rewriteAs} \ear\no 
where 

\be
s = {x_{14}^2x_{23}^2\over x_{12}^2x_{34}^2}\ , \qquad t = 
{x_{13}^2x_{24}^2\over x_{12}^2x_{34}^2}\ . \label{defst} \ee\no 
The entire G-analytic superfield four-point function can then be 
constructed systematically as a power series in $\th^+$ and 
$\bar\th^+$ using the superconformal Ward identities 
\cite{announced}. In addition, the dynamical requirement of 
H-analyticity at the  $\theta^+\bar\theta^+$ level leads to the 
following differential constraints on the coefficient functions 
$a,b,c$: \footnote{The result we quote here follows from the 
general properties of the correlation functions and should be 
valid to all orders in perturbation theory, as will be shown in 
detail elsewhere \cite{announced}.} 

\bear \beta_s &=& -t\alpha_s -t\alpha_t\ , \non\\ \beta_t &=& 
s\alpha_s  + (s-1)\alpha_t\non\\ \label{diffconst} \ear\no where 
$\alpha,\beta$ are determined from the linear equations 

\bear b &=& \alpha + {\beta\over t} + {a\over s}\ , \non\\ c &=& - 
{s\over t}\beta + {t-s-1\over s}a \ . \non\\ 
\label{determalphabeta} \ear\no One sees that only two of the 
three coefficients are restricted. The set of first-order partial 
differential equations (\ref{diffconst}) is equivalent to a single 
second-order equation (the compatibility condition for the system 
(\ref{diffconst})), e.g., for $\alpha$: 

\be
s\alpha_{ss} + t\alpha_{tt}+(s+t-1)\alpha_{st}
+2(\alpha_s+\alpha_t) = 0 \, .
\label{eqalpha}
\ee\no
Given some boundary conditions the above equations allow one to
completely fix $\alpha$ and $\beta$, thus leaving arbitrary only the
coefficient $a$. 

In this note we shall concentrate on applying this general 
relation at the two-loop level. Here we have some additional 
information which simplifies the partial differential equations 
(\ref{diffconst}). From the explicit graph calculation we already 
know that (see (\ref{resA1}),(\ref{rewriteAs})) 

\be
a = 4\pi^4 s\Phi^{(1)}(s,t),\qquad b = 4\pi^4\Phi^{(1)}(s,t).
\label{ab=phi}
\ee\no
In terms of $\alpha$ and $\beta$ this becomes
$\beta = -t\alpha$, after which the equations (\ref{diffconst})
can be easily solved to yield the following general solution:

\be
\alpha = {\mu\over s}\ ,\qquad \beta = -\mu {t\over s}\ , \qquad 
\mu = \mbox{const}\ . \label{solvediffconst} \ee\no Combining eqs. 
(\ref{rewriteAs}), (\ref{determalphabeta}), (\ref{ab=phi}) and 
(\ref{solvediffconst}) we find 

\be
x_{12}^2x_{23}^2x_{34}^2x_{14}^2
A_3
=
\mu -4\pi^4 (s-t+1)\Phi^{(1)}(s,t). \label{relA3Phi} \ee\no The 
constant $\mu$ is easily shown to be zero using the known 
behaviour \cite{ehssw} of $A_3$ and $\Phi^{(1)}$ in the limit 
$x_{14}\to 0$. Thus we arrive at the following expression for 
$A_3$: 

\be
A_3
=
4\pi^4 {x_{13}^2x_{24}^2-x_{14}^2x_{23}^2-x_{12}^2x_{34}^2\over 
x_{12}^4x_{34}^4 x_{23}^2 x_{14}^2} \,\Phi^{(1)} \biggl( 
{x_{14}^2x_{23}^2\over x_{12}^2x_{34}^2}, {x_{13}^2x_{24}^2\over 
x_{12}^2x_{34}^2}\biggr)\ . \label{realA3Phifin} \ee\no By 
(\ref{A3}) this result can be expressed as a partial differential 
equation for $f$, \bear x_{12}^2x_{34}^2 (\partial_2 + 
\partial_3)^2 f(x_1,x_2,x_3,x_4) + ( 1 \leftrightarrow 3 ) &=& 
\non\\ &&\hspace{-110pt} 4\pi^4\Biggl\lbrack 
{x_{13}^2x_{24}^2-x_{14}^2x_{23}^2-x_{12}^2x_{34}^2\over 
x_{12}^2x_{34}^2} \,\Phi^{(1)} \biggl( {x_{14}^2x_{23}^2\over 
x_{12}^2x_{34}^2}, {x_{13}^2x_{24}^2\over x_{12}^2x_{34}^2}\biggr) 
\non\\ && \hspace{-110pt} + {x_{12}^2+x_{14}^2-x_{24}^2\over 
x_{24}^2} \,\Phi^{(1)}\Bigl( {x_{12}^2\over 
x_{24}^2},{x_{14}^2\over x_{24}^2}\Bigr) + 
{x_{12}^2+x_{23}^2-x_{13}^2\over x_{13}^2} \,\Phi^{(1)}\Bigl( 
{x_{12}^2\over x_{13}^2},{x_{23}^2\over x_{13}^2}\Bigr) \non\\ && 
\hspace{-110pt} +{x_{23}^2+x_{34}^2-x_{24}^2\over x_{24}^2} 
\,\Phi^{(1)}\Bigl({x_{23}^2\over x_{24}^2},{x_{34}^2\over 
x_{24}^2} \Bigr) + {x_{14}^2+x_{34}^2-x_{13}^2\over x_{13}^2} 
\,\Phi^{(1)}\Bigl( {x_{14}^2\over x_{13}^2}, {x_{34}^2\over 
x_{13}^2}\Bigr) \Biggr\rbrack\ . \non\\ \label{idnontriv} \ear 
This equation is non-trivial, and we have not been able to show 
that it is satisfied analytically, since $f$ itself is not known 
in sufficiently explicit form. Instead, we have used MATHEMATICA 
to evaluate both sides of the equation numerically, and  have 
verified that they agree for more than twenty choices of the 
parameters $x_{ij}^2$ (with three-digit precision). For the 
evaluation of $f$ we used a two-parameter integral representation 
\cite{davydychevpriv} which relates the integral (\ref{2loop}) to 
the sunset diagram with one massless line 
\cite{generalis,djouadi,tausk,bertau}. On the right hand side the 
explicit formula (\ref{Phi1explicit}) was employed. 

Retrospectively, the knowledge that all three coefficient 
functions in (\ref{answer}) are related to each other, so that it 
is sufficient to determine one of them, can be used to drastically 
simplify the graph calculation carried out in \cite{ehssw}. 
Suppose that we wish to compute only the coefficient $A_2$. To 
this end we can identify two pairs of harmonic variables as 
follows, 

\be
u_1\equiv u_4\ ,\qquad u_2 \equiv u_3\ , \label{identharm} \ee\no 
so that only the structure $(12)^2(34)^2\to (12)^4$ survives. Then 
among all the two-loop graphs considered in \cite{ehssw} only the 
following one is non-vanishing: 

\vskip 0.5 in
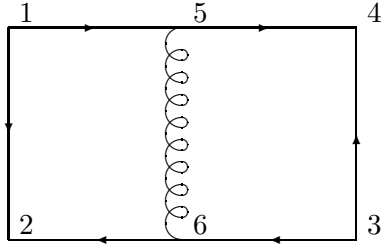
\begin{center}
\begin{picture}(18000,12000)(0,-4000)

\startphantom
\drawline\gluon[\E\CENTRAL](0,0)[14]
\stopphantom
\global\Xone=\gluonlengthx
\global\Xeight=\Xone
\global\divide\Xeight by 4
\drawline\fermion[\E\REG](0,0)[\Xone]
\drawarrow[\W\ATTIP](\Xeight,0)
\global\multiply\Xeight by 3
\drawarrow[\W\ATTIP](\Xeight,0)
\put(400,200){2}
\global\advance\pmidx by -400
\global\Ytwo=-1500
\global\advance\pmidx by 800
\put(\pmidx,200){6}
\global\advance\pmidx by -400
\global\advance\pbackx by 400
\put(\pbackx,200){3}
\drawline\gluon[\N\CENTRAL](\pmidx,\pmidy)[8]
\global\advance\pmidx by -300
\global\advance\pmidy by 270
%\drawarrow[\SE\ATTIP](\pmidx,\pmidy)
\global\Yone=\gluonlengthy
\drawarrow[\E\ATTIP](\Xeight,\Yone)
\global\divide\Xeight by 3
\drawarrow[\E\ATTIP](\Xeight,\Yone)
\global\advance\Yone by 200
\global\advance\pbackx by 400
\put(\pbackx,\Yone){5}
\global\divide\Xone by 2
\drawline\fermion[\W\REG](\gluonbackx,\gluonbacky)[\Xone]
\global\advance\pbackx by 400
\put(\pbackx,\Yone){1}
\drawline\fermion[\E\REG](\gluonbackx,\gluonbacky)[\Xone]
\global\advance\pbackx by 400
\put(\pbackx,\Yone){4}
\global\advance\Yone by -200
\drawline\fermion[\S\REG](\fermionbackx,\pbacky)[\Yone]
\drawarrow[\N\ATTIP](\pmidx,\pmidy)
\drawline\fermion[\N\REG](0,0)[\Yone]
\drawarrow[\S\ATTIP](\pmidx,\pmidy)

%\put(4800,-4000){Figure 1}
\end{picture}
\end{center}

We recall that the matter propagators contain eight spinor derivatives,
e.g.

\vskip 0.5 in
\begin{center}
\begin{picture}(30000,4000)(0,-2000)
\put(0,2000){\scriptsize{1}} 
\drawline\fermion[\E\REG](1500,2000)[6430] 
\drawarrow[\E\ATTIP](\pmidx,\pmidy) 
\put(8430,2000){\scriptsize{5}} \put(15000,2000){$ 
{(D^+_1)^4(D^+_5)^4\delta^8(\theta_1-\theta_5)\over 
(15)^3x_{15}^2} $} 
%\put(1500,-1500){Figure 2}
\end{picture}
\end{center}
whereas the SYM one has only four:

\vskip 0.5 in
\begin{center}
\begin{picture}(30000,4000)(0,-2000)
\put(0,2000){\scriptsize{5}}
\drawline\gluon[\E\REG](1500,2000)[6]
\put(8430,2000){\scriptsize{6}}
\put(15000,2000){$
{(D^+_5)^4\delta^8(\theta_5-\theta_6)\over x_{56}^2}
\delta(u_5,u_6)$}
%\put(1500,-1500){Figure 1}
\end{picture}
\end{center}
The vertices $5$ and $6$ are G-analytic, i.e. they involve the  
Grassmann integral measures $\int  d^4\theta^{+}$. As usual in 
supergraph calculations, these measures have to be restored to the 
full superspace ones according to the identity $\int 
d^4\theta^{+}(D^{+})^4 = \int d^8\theta$. At the vertices $5$ and 
$6$ this can be done with the help of the spinor derivatives 
$(D^{+}_5)^4$ and $(D^{+}_6)^4$ from, e.g., the propagators $1-5$ 
and $6-2$. The remaining spinor derivatives $(D_1^+)^4$ and 
$(D_2^+)^4$ can then be integrated by parts onto the SYM line or 
the other two matter lines $5-4$ and $3-6$. However, owing to the 
identification (\ref{identharm}), the propagator $5-4$ has the 
same G-analyticity as $1-5$, and similarly for $3-6$ and $6-2$. 
Thus, all the spinor derivatives from $1-5$ and $6-2$ go onto the 
SYM line and we get twelve spinor derivatives on 
$\delta^8(\theta_5-\theta_6)$. Next, since the propagators $1-5$ 
and $6-2$ are now spinor derivative-free, the $\theta$ points $5$ 
and $6$ are identified with the external ones $1$ and $2$. 
Finally, we are only interested in the leading (scalar) term of 
the amplitude, so we can set $\theta_{1,2,3,4}=0$. This results in 
a D'Alembertian on the internal line $5-6$, which explains why the 
two-loop integral collapses to the one-loop integral $\Phi^{(1)}$. 

To summarise, we have shown that, by combining the results of 
\cite{ehssw} with the constraints implied by superconformal 
invariance and H-analyticity, we are able to express the 
coefficient function $A_3$, which seemed to involve a ``true'' 
two-loop integration, and had so far defied all efforts at 
analytic evaluation, in terms of the functions $A_1,A_2$ which are 
easily expressed in terms of a single one-loop integral. This 
relation between $A_1,\ A_2$ and $A_3$ implies a non-trivial 
partial differential equation for the two-loop integral $f$ which 
we have confirmed numerically. We have also demonstrated that, 
once it is known that the calculation of the complete amplitude 
can be reduced to the calculation of either $A_1$ or $A_2$, this 
knowledge can be used to drastically simplify the Feynman diagram 
calculation. 

We should mention that neither the two-loop calculation performed 
in \cite{ehssw} nor the analysis of the Ward identities presented 
here make any use of the full $N=4$ supersymmetry. In addition to 
$N=2$ supersymmetry, they rely only on finiteness (conformal 
invariance) which is known to be a property of a larger class of 
supersymmetric theories. At present we cannot say what specific 
r\^ole $N=4$ supersymmetry might play at higher orders of 
perturbation theory. 

It should be emphasized that our final result is much simpler than 
one would a priori expect. Generically, a four-point two-loop 
integral such as the one encountered here is expected to contain, 
in addition to logarithms and dilogarithms of the kinematic 
variables, also tri- and quadrilogarithms. And indeed, the elusive 
function $f$ itself contains polylogarithms up to fourth order, as 
can be seen from the limiting case $x_{23}=0$, for which $f$ was 
obtained in closed form in \cite{davuss2loop3point}. It is 
tempting to speculate that the absence of tri- and 
quadrilogarithms in the two-loop result is just a glimpse of a 
universality property which would extend to all orders in 
perturbation theory for this amplitude, at least for large $N_c$. 
This seems even more plausible in view of the recent result that 
the same one-loop box function $\Phi^{(1)}$ appears also in the 
instanton contribution to four-point correlators in $N=4$ SYM 
theory \cite{bgkr}. The same simple dilogarithmic behaviour is 
found also for the spacetime dependence of the axion/dilaton 
amplitudes corresponding to the $N=4$ four-point function 
according to the Maldacena conjecture \cite{hong,dzf,dzf2,dfmmr}.

\vspace{20pt} {\bf Acknowledgements:} We are indebted to A. 
Davydychev for repeatedly sharing with us his expertise on 
two-loop massless scalar integrals. We would also like to thank Z. 
Bern, O. Dor\'e and K. Intriligator for helpful discussions, and 
D. Fliegner for computer support. This work was supported in part 
by the EU network on Integrability, non-perturbative effects, and 
symmetry in quantum field theory (FMRX-CT96-0012) and by the 
British-French scientific programme Alliance (project 98074).

\end{document}